\documentclass[aps,pra,twocolumn,preprintnumbers,amsmath,amssymb]{revtex4}
\usepackage{graphicx}
\usepackage{dcolumn}
\usepackage{bm}
\usepackage{paralist}
\usepackage{color}

\newcommand{\ud}{\mathrm{d}}

\newcommand{\mb}[1]{\boldsymbol{#1}}
\newcommand{\br}{\mb{r}}
\newcommand{\bx}{\mb{x}}

\usepackage{slashed}

\newcommand{\braket}[2]{\langle{#1}|{#2}\rangle}

\newcommand{\abrakket}[3]{\left\langle {#1}\middle|{#2}\middle|{#3} \right\rangle}
\newcommand{\brakket}[3]{\langle{#1}|{#2}|{#3}\rangle}

\newlength{\back}

\allowdisplaybreaks[2]

\begin{document}

\title{Asymptotic behavior of the electron density and the Kohn-Sham potential in case of a Kohn-Sham HOMO nodal plane} 
\author{Paola Gori-Giorgi$^1$}
\author{Tam\'as G\'al$^{1,2}$}
\author{Evert Jan Baerends$^{1,2}$}
\affiliation{$^1$Theoretical Chemistry, VU University, De Boelelaan 1083, 1081 HV Amsterdam, The Netherlands\\
$^2$WCU program, Pohang University of Science and Technology,  Pohang 790-784, South-Korea}

\date{\today}

\begin{abstract}
It is known that the asymptotic decay of the electron density $n(\br)$ outside a molecule is informative about its first ionization potential $I_0$, $n(|\br|\to\infty) \sim \text{exp}(-2\sqrt{2I_0}\,r)$. This dictates the orbital energy of the highest occupied Kohn-Sham (KS) molecular orbital (HOMO) to be $\epsilon_H=-I_0$, if the KS potential goes to zero at infinity.  However, when  the Kohn-Sham HOMO has a nodal plane, the KS density in that plane will  decay as $\exp{(-2\sqrt{-2\epsilon_{H-1}}\,r)}$. Conflicting proposals exist for the KS potential: from exact exchange calculations it has been found that the KS potential approaches a {\em positive} constant in the plane, but from the assumption of isotropic decay of the exact (interacting) density it has been concluded this constant needs to be {\em negative}. Here we show that either 1) the exact density decays differently (according to the second ionization potential $I_1$) in the HOMO nodal plane than elsewhere, and the KS potential has a regular asymptotic behavior (going to zero everywhere) provided that $\epsilon_{H-1}=-I_1$; or 2) the density does decay like $\text{exp}(-2\sqrt{2I_0}\,r)$ everywhere but the KS potential exhibits strongly irregular if not divergent behavior around (at) the nodal plane.
\end{abstract}

\maketitle
\section{Introduction}
In Kohn-Sham (KS) density functional theory (DFT), one of the most widely used techniques in electronic structure theory, ground-state properties are calculated via the KS system, consisting of non-interacting electrons moving in the local KS potential $v_s(\br)$ \cite{KohnSham1965a}. In principle, the KS potential $v_s(\br)$ ensures that the electron density $n(\br)$ of the non-interacting KS system is the same as that of the physical, interacting system. Exact properties of $n(\br)$ and $v_s(\br)$ have played -- and continue to play -- a crucial role in constructing and improving approximations.
 
Both the square root of the density and the KS orbitals $\psi_k(\br)$ obey Schr\"odinger-type equations, 
\begin{eqnarray}
\label{eq:densityeqn}
\left(-\frac{1}{2}\nabla^2+v_{\rm ext}(\br)+v_{\rm eff}(\br)\right)\sqrt{n(\br)}  =   -I_0\sqrt{n(\br)}  \\
\label{eq:KSeqn}
\left(-\frac{1}{2}\nabla^2+v_{\rm ext}(\br)+v_{\rm Hxc}(\br)\right)\psi_k(\br)  = \epsilon_k\psi_k(\br),
\end{eqnarray}
where the sum of the external and the Hartree-exchange-correlation potentials constitutes the KS potential, $v_s=v_{\rm ext}+v_{\rm Hxc}$. The eigenvalue in Eq.~\eqref{eq:densityeqn} \cite{Hunter1975Symp9,LevyPerdewSahni1984} is the first ionization potential, $I_0=E_0^{N-1}-E_0^N$, and the occupied KS orbitals reproduce the density, $\sum_{k}^N|\psi_k(\br)|^2=n(\br)$.

Here we are mainly concerned with molecules, where the external potential $v_{\rm ext}(\br)$ goes to zero at large distance like $-Z/r$, with $Z$ representing the total charge of all nuclei and $r$ the distance from the center of nuclear charge. In this case, according to Eqs.~(\ref{eq:densityeqn})-(\ref{eq:KSeqn}), the asymptotic ($r\to\infty$) decay of $\sqrt{n(\br)}$ and $\psi_k(\br)$ is  
\begin{eqnarray}
\sqrt{n(\br)}	& \sim &  e^{-\sqrt{2(I_0+v_{\rm eff}(\infty))}\,r} \\
 \psi_k(\br) &	\sim & e^{-\sqrt{2(-\epsilon_k+v_{\rm Hxc}(\infty))}\,r}.
\end{eqnarray} 
Both the effective potential $v_{\rm eff}(\br)$ for $\sqrt{n(\br)}$ and the Hartree-exchange-correlation potential $v_{\rm Hxc}(\br)$ had been thought, until recently, to go to zero asymptotically everywhere in space (in other words, it seemed always possible to fix the arbitrary constant in the functional derivatives for finite systems such that the effective potentials go to zero in all possible directions). In the exact KS model (and in exact generalized KS models as well \cite{SeidlGorlingVoglMaiewski1996}) the model density, which decays as the square of the highest occupied molecular orbital, should decay like the exact density, leading to the identification $I_0=-\epsilon_H$ \cite{LevyPerdewSahni1984,AlmbladhBarth1985}. 

In case the KS HOMO has a nodal plane (HNP) extending to infinity, a very straightforward argument was given by Wu {\it et al.}~\cite{WuAyersYang2003} that the KS potential should go to a {\em negative} constant for asymptotic points $r_p \to \infty$ in that plane. 
The KS density in that plane is governed by the HOMO$-1$, assuming HOMO$-1$ does not have the same nodal plane. Wu {\it et al.}~\cite{WuAyersYang2003} made the common assumption that the exact interacting density has the same asymptotic behavior everywhere, and they observed  that then the decay of the HOMO$-1$ in the HNP must be equal to the decay of the total density, implying $-\sqrt{2(-\epsilon_{H-1}+v_{\rm Hxc}(r_p \to \infty))}\,r_p=-\sqrt{2I_0}\,r_p$, so that $v_{\rm Hxc}(r_p \to \infty)$ should tend to the {\em negative} constant $I_0+\epsilon_{H-1}=-(\epsilon_H-\epsilon_{H-1})$. 
On the other hand, it had earlier been argued, and numerical evidence had been provided, that the optimized effective potential method for the exact exchange model (xOEP) of Kohn-Sham theory leads to an asymptotic constant in the HNP, but {\em positive}~\cite{DellaSalaGoerling2002,DellaSalaGoerling2002b,KuemmelPerdew2003,KuemmelPerdew2003b}. Since it was pointed out that this behavior of the Kohn-Sham potential has a significant effect on the orbital energies of particularly the higher lying unoccupied orbitals, with large consequences for the excitation energies calculated with time-dependent DFT \cite{DellaSalaGoerling2002,DellaSalaGoerling2002b}, the matter is also relevant for practical calculations.

In this work we analyze this issue, starting from a very simple question: is it true that an asymptotic constant on the HNP in the KS potential changes the exponential decay of the HOMO-1 (and of all the other orbitals) on that plane? We will see in Sec.~\ref{sec:angularlaplacian} that the argument of Wu {\it et al.}~\cite{WuAyersYang2003} did not consider the role of the angular part of the laplacian, which, instead, cannot be neglected.  

We then turn to the key question: how does the exact density behave? We use (see section \ref{Sec_asymptoticn}) the expansion of the density in the squares of the Dyson orbitals. We analyze the energy-independent equations obeyed by the Dyson orbitals, in order to derive the behavior of the two leading Dyson orbitals, the first with eigenvalue $-I_0$ and the second with eigenvalue $-I_1$,  and of the density in and close to the plane. In section \ref{sec:Case1} we discuss the case (Case 1) in which the KS potential can have a simple and regular behavior (the expected isotropic $-1/r$ asymptotics everywhere). In this case it is necessary that the asymptotic decay of the exact interacting density be different in different directions: in the  HOMO nodal plane the density decay $n(|\br_p|\to\infty)$ should be faster (according to the second ionization potential, $\text{exp}[-2\sqrt{2I_1}\,r_p]$) than  outside the plane, and the KS orbital energy of the HOMO$-1$ should be equal to the second ionization potential, $-\epsilon_{H-1}=I_1$. It depends on the properties of the system (e.g. spatial and spin symmetries of the ion states) if such a situation will occur, with for instance the prototype molecules ethylene and benzene \cite{DellaSalaGoerling2002b} as good candidates.
However, such behavior of the density  will not always occur. In section \ref{sec:Case2} we discuss the case (Case 2) that the density has the same exponential decay, governed by $I_0=-\epsilon_H$, on the nodal plane as everywhere else (although it must then decay \textit{polynomially} slower). In that case the KS potential cannot decay uniformly like $-1/r$ but has to exhibit very irregular behavior near the plane (possibly even divergent behavior in the plane) in order to impart on the KS orbitals the shapes that will make them reproduce the true density.

\section{Asymptotic decay of the orbitals on a plane and their corresponding potential}
\label{sec:angularlaplacian}
The usual argument to derive the asymptotic decay of the KS orbitals for a finite system (or, in general, of a single-particle Schr\"odinger equation with a multiplicative potential), is based on the fact that if the potential $v_s(\br)$ goes to 0 when $|\br|\to\infty$, then asymptotically the single particle equation reads $-\frac{1}{2}\nabla^2 \psi=\epsilon\, \psi$ implying $\psi(\br)\sim e^{-\sqrt{-2 \epsilon}\,r}$. If, instead, the potential goes asymptotically to a constant $v_s(\infty)$, then we trivially obtain from the same equation $\psi(\br)\sim e^{-\sqrt{2(-\epsilon+v_s(\infty))}\,r}$. This leads to the idea that if the exponential decay of $\psi(\br)$ is the same everywhere except on a plane, then in that plane the potential has to go asymptotically to a constant. And viceversa, one expects that an asymptotic constant in $v_s(\br)$ on a plane implies that the asymptotic decay of $\psi(\br)$ on the plane has a different exponent. 

This argument, however, does not take into account the fact that the laplacian also contains angular derivatives. This angular part has a $1/r^2$ prefactor and one may think that this makes it negligible, when $r\to\infty$, with respect to the constant terms (the eigenvalue and, if present, the constant $v_s(\infty)$). However, if we have a different exponential decay on the plane than elsewhere, the relative difference between the orbital on the plane and very close to it increases exponentially with $r$, so that the angular derivative of the orbital very close to the plane increases much faster than $1/r^2$. This diverging behavior needs to be compensated by $v_s(\br)$. We illustrate this with two very simple examples.

\begin{figure}
   \includegraphics[width=7cm]{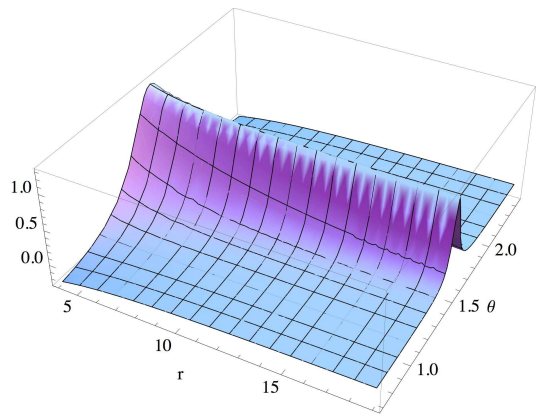}
   \caption{The large-$r$ behavior of the potential that generates the orbital with the asymptotic behavior of Eq.~\eqref{eq:phi1} as a function of $r$ and the azimuthal angle $\theta={\rm arccos}(z/r)$. We clearly see that this potential has a ``ridge'' on the $xy$ plane (corresponding to $\theta=\pi/2$), where it goes asymptotically to a constant. The ``ridge'' shrinks as $r$ gets larger and larger. Despite this constant, the orbital has the same exponential decay everywhere, it only decays $1/r^2$ faster on the plane.}
\label{fig_example1}
\end{figure}

First of all, let us consider an example that shows that an asymptotic constant on a plane in the potential does not necessarily imply  a change in the exponential decay of the orbital on the plane. Suppose we have an orbital with the following asymptotic behavior
\begin{equation}
\psi(\br)\sim e^{-r}\left(\cos(\theta)^2+\frac{1}{r^2}\right),
\label{eq:phi1}
\end{equation}
where $\theta=\pi/2$ defines the $xy$ plane in spherical coordinates. This orbital has the same asymptotic exponential decay everywhere, but on the $xy$ plane it decays $1/r^2$ faster than elsewhere. We can compute the corresponding potential by inversion, $v(\br)=\nabla^2\psi({\br})/2\psi(\br)$, and we find that $v(r\to\infty,\theta)=1/2$, but $v(r\to\infty,\pi/2)=3/2$. The potential is shown in Fig.~\ref{fig_example1} as a function of $r$ and $\theta$ (we have subtracted 1/2 so that $v(\br)$ goes asymptotically to zero outside the plane). We clearly see that this potential has a ``ridge'' on the $xy$ plane, where it goes asymptotically to a constant. The ``ridge'' shrinks as $r$ gets larger and larger. This kind of behavior in the potential has been usually associated in the literature to a change in the exponential decay of the orbital on the plane, but we see here a clear counterexample.

\begin{figure}
   \includegraphics[width=7cm]{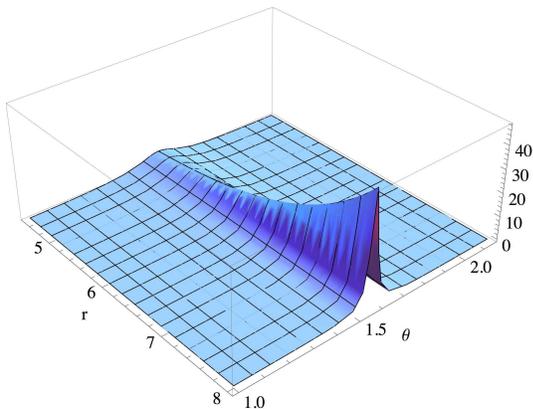}
   \caption{The large-$r$ behavior of the potential that generates the orbital with the asymptotic behavior of Eq.~\eqref{eq:phi2} as a function of $r$ and the azimuthal angle $\theta={\rm arccos}(z/r)$. We clearly see that this potential has a ``ridge'' on the $xy$ plane (corresponding to $\theta=\pi/2$), where it diverges exponentially. The ``ridge'' shrinks as $r$ gets larger and larger. }
\label{fig_example2}
\end{figure}

As a second example, we consider an orbital with an exponentially faster decay on the plane:
\begin{equation}
\psi(\br)\sim e^{-r}\cos(\theta)^2+e^{-2 r}.
\label{eq:phi2}
\end{equation}
Again, we compute the corresponding potential by inversion, and we find that this potential has also a ``ridge'' on the plane, where it diverges exponentially, as shown in Fig.~\ref{fig_example2}. Again, the ``ridge'' shrinks as $r\to\infty$. This example shows that a different exponential decay of the orbital on the plane does not necessarily imply that the corresponding potential goes to a constant on the plane: as we see here the potential might diverge, in order to compensate the derivative perpendicular to the plane, which increases exponentially as $r$ increases. 
In section \ref{sec:Case2} we will discuss the case that there is $slower$ decay of an orbital on the plane than elsewhere, where it will be demonstrated that dependent on how this slower decay on the plane is approached ($\theta$-dependence), the potential may exhibit asymptotically either a negative constant, or may diverge to $+\infty$ or to $-\infty$. Even in the case of a negative constant on the plane, we will see that the potential close to the plane displays very irregular behavior.

\section{Asymptotic behavior of the exact density from the Dyson orbital expansion}
\label{Sec_asymptoticn}
Although it is commonly assumed in the literature on the asymptotic decay of the density that the exact density decays everywhere in the same way, it should be recognized that this is not always true. First of all, it is evident for a noninteracting electron system, that if there is a nodal plane in the HOMO, the density decays differently in that plane, because the HOMO density does not contribute in the plane.   
However, the interacting case, where a configuration mixing will involve many configurations, is more subtle.  While in a non-interacting system described by a single determinant  a HOMO nodal plane determines a nodal plane for the whole wavefunction (the wavefunction when all the electrons are on the nodal plane is zero), such a nodal plane in general does not survive in the many-configuration interacting wavefunction. It is also known that in the case of nodes in the wavefunction due to the fermionic character of the electrons (antisymmetry of the wavefunction under permutation) electron interaction substantially modifies the nodes \cite{Ceperley1991,Mitas2006}.

To study the density decay in the general interacting case, we express the exact $N$-electron wavefunction $\Psi^N_0$ and the exact density in terms of the Dyson orbitals $d_i(\bx)$,
\begin{eqnarray}
\Psi^N_0 &=& N^{-1/2}\sum_{i=0}^\infty d_i(\bx) \Psi^{N-1}_i(2 \cdots N), \nonumber  \\
d_i(\bx) & = & \sqrt{N} \int{ \Psi^{N-1}_i(2 \cdots N)^* \Psi^N_0(\bx,2 \cdots N)} \ud 2 \cdots \ud N,  \nonumber \\
n(\bx) & = & \sum_{i=0}^{\infty} |d_i(\bx)|^2,  \label{eq:Dysonorb}
\end{eqnarray}
where the $\Psi^{N-1}_i$ are the exact $(N-1)$-electron states and $\bx = \br,s$. The sum over $i$ goes over both the spin-$\uparrow$ and spin-$\downarrow$ Dyson orbitals. If e.g. $\bx=(\br,s=\uparrow)$ then only the spin-$\uparrow$ Dyson orbitals are nonzero at $\bx$ and contribute to $n(\bx)=n(\br,\uparrow)$ ($=\frac{1}{2}n(\br)$ in closed shell systems).
Each state of the ion is associated with a one-particle wavefunction, its Dyson orbital. These orbitals constitute a nonorthogonal nonnormal, in general linearly dependent set. We define the conditional amplitude $\Phi(2 \cdots N;\bx)$ \cite{Hunter1975} and associated quantities,
\begin{eqnarray}
\label{eq:condampl}
\Phi(2 \cdots N;\bx) & = & \frac{\Psi^N_0(\bx,2 \cdots N)} {\sqrt{n(\bx)/N} }, \nonumber \\
n^{cond}(\bx_2|\bx) & = & (N-1)\int |\Phi(2\cdots N|\bx)|^2 \ud 3\cdots \ud N, \nonumber  \\
v^{cond}(\bx) & =& \int{ \frac{n^{cond}(\bx_2|\bx)}{|\br-\br_2|} } \ud \bx_2.
\end{eqnarray}
$\Phi(2 \cdots N;\bx)$ is a normalized $(N-1)$-electron wavefunction depending parametrically on the position $\bx$. Its square describes the probability distribution of electrons at positions $2 \cdots N$ when one electron is known to be at $\bx$. Its associated one-electron density $n^{cond}(\bx_2|\bx)$ is the density of the other electrons at position $\bx_2$ when one electron is at $\bx$, which is the normal one-electron density $n(\bx_2)$ plus the full exchange-correlation hole surrounding position $\bx$, $n^{cond}(\bx_2|\bx)=n(\bx_2)+n_{xc}^{hole}(\bx_2|\bx)$, $v^{cond}(\bx)=\int d\bx_2 n^{cond}(\bx_2|\bx)/|\bx_2-\bx| = v_{Hartree}(\bx)+v_{xc}^{hole}(\bx)$. 
Projecting the Schr\"odinger equation $\hat{H}^N \Psi^N_0=E_0^N \Psi_0^N$ against $\Psi_i^{N-1}(2 \cdots N)$ and using the expansion of Eq.~\eqref{eq:Dysonorb} one obtains the energy-independent equations for the Dyson orbitals,
\begin{equation}
\label{eq:Dysoneqn}
\left(-\frac{1}{2}\nabla^2+v_{\rm ext}(\br)\right)d_i(\bx) +\sum_{k=0}^\infty X_{ik}(\bx)d_k(\bx) =  -I_id_i(\bx).
\end{equation}
Katriel and Davidson (KD) \cite{KatrielDavidson1980} pointed out that, due to the coupling integrals
\begin{equation}
  X_{ik}(\bx)\equiv\brakket{\Psi_i^{N-1}}{\sum_{j > 1}^N\frac{1}{|\br_j-\br|} }{\Psi_k^{N-1}}_{2..N},
\end{equation} 
the exponential decay of the coupled Dyson orbitals will be the same. This was demonstrated by Handy {\it et al.}~\cite{HandyMarronSilverstone1969} for the analogous case of coupling of the Hartree-Fock orbitals by the exchange term. The first Dyson orbital $d_0$ will have exponential decay $\sim e^{-\sqrt{2I_0}\,r}$ multiplied by a factor $r^{\beta}$ with $\beta=(Z-N+1)/\sqrt{2I_0}-1$, due to the $-Z/r$ decay of $v_{\rm ext}$ and the $(N-1)/r$ decay of the coupling term. KD find that higher Dyson orbitals which have nonzero $X_{i0}$ with the first Dyson orbital will have decay $r^{\beta-L^*}e^{-\sqrt{2I_0}\,r}$ with $L^* \ge 2$. Dyson orbitals that are not connected to  $d_0$ will have different exponential  decay, governed by the eigenvalue of the first orbital in such a connected set (which is disjunct from other sets).
Considering the expansion of the density in Dyson orbitals in Eq.~\eqref{eq:Dysonorb}, KD have concluded that, if the density decays for $|\br| \to \infty$ as the most slowly decaying term $|d_0(\bx)|^2$, its exponential decay would be $\sim e^{-2\sqrt{2I_0}\,r}$. Levy, Perdew and Sahni (LPS) \cite{LevyPerdewSahni1984} proved this exponential decay in a different way, thereby proving that the leading term is not overruled by the infinite sum of the faster decaying terms in Eq.~\eqref{eq:Dysonorb}. The result 
\begin{equation}
n(|\br|\to\infty) \sim |d_0(\br)|^2 \sim e^{-2\sqrt{2I_0}\,r} 
\end{equation}
is considered well established.

This picture changes if the KS HOMO has a nodal plane. The common thinking is that always the interacting density decays everywhere in the same way due to correlation effects  \cite{WuAyersYang2003,Holas2008}. Instead, by analyzing the Dyson orbitals we can see that a KS HNP may also imply a special behavior of the interacting density. A nodal plane extending to infinity is typically related to a symmetry plane (consider, e.g., ethylene or benzene \cite{DellaSalaGoerling2002,KuemmelPerdew2003}), but a HOMO nodal surface can occur also in more general situations.
In the case of a symmetry plane, the exact interacting states of the molecule are either symmetric or antisymmetric with respect to the plane. For example, the ground state wavefunction corresponding to a closed shell configuration is totally symmetric with respect to that plane, while the first ion state $\Psi_0^{N-1}$ will be antisymmetric (the KS first ion state surely will be so, and  we will consider the usual case that the same holds for the exact ion state). For points $\br_p$ in the HNP the conditional amplitude $\Phi(2 \cdots N|\bx_p)$ will be symmetric with respect to the plane. Therefore, the matrix element  $\braket{\Psi_0^{N-1}}{\Phi(2 \cdots N|\bx_p)}_{2..N}$ will vanish, so that the first Dyson orbital is zero in the plane:
\begin{align}
\label{d0} 
&d_0(\bx_p)=\sqrt{n(\bx_p)} \braket{\Psi_0^{N-1}}{\Phi(2 \cdots N|\bx_p)}_{2..N}=0. 
\end{align}
In fact, $d_0$ is antisymmetric with respect to the plane.  To obtain the asymptotic behavior of higher Dyson orbitals in the plane, we have to solve Eq.~\eqref{eq:Dysoneqn} for points $\bx_p$ in the plane. Since  $d_0(\bx_p)=0$, it looks as if the coupling to $d_0$  will be absent for any higher Dyson orbital $d_{i>0}$. The decay in the HNP of the second Dyson orbital (and, thus, of the density) is then not governed by $d_0$, but will be according to the second ionisation potential, $I_1$ (if $d_1$ is not also zero on the plane).  We will in section \ref{sec:Case1} discuss this situation (Case 1), which would allow for a smooth Kohn-Sham potential that decays uniformly as $-1/r$.  However, a closer scrutiny of Eq.~\eqref{eq:Dysoneqn} reveals that in general $d_1$ and the density will inherit in the HNP the slower decay according to $I_0$ from $d_0$. In the KS case the HOMO$-1$  does not couple to another orbital, and the asymptotic behavior of the density in the HNP is determined by the HOMO$-1$. Given its eigenvalue close to $-I_1$, and the corresponding ``fast" decay everywhere else, the slow decay according to $I_0$ on only the HNP  requires very irregular features in the Kohn-Sham potential, as will be discussed in section \ref{sec:Case2}.

\section{Case 1: The density decay on the HNP is governed by the second ionization potential}\label{sec:Case1}

\subsection{Asymptotic behavior of the density}\label{sec:Case1asymptoticdensity}
Since $d_0=0$ in the HNP, we have to turn to  $d_1(\bx_p)$ to determine the asymptotic behavior of the density on the plane, $n(|\br_p| \to \infty)$. If the corresponding excited ion state $\Psi_1^{N-1}$ has the same symmetry with respect to the HNP as $\Psi_0^N$, this Dyson orbital will not be zero in the plane (see Eq.\ \eqref{eq:Dysonorb}).  
In this section we explore the case that  there would be no coupling between $d_1$ and $d_0$, which would be the case if  $X_{10} \equiv 0$.  For simplicity we only consider the coupling of $d_1$ to $d_0$ and consider the asymptotic terms in the equation for $d_1$
\begin{align}\label{eq:d1}
\left( -\frac{1}{2}\nabla^2+v_{ext} \right) d_1+X_{11}d_1  + X_{10}d_0=-I_1d_1  
\end{align}    
The term with $X_{11}$ is not problematic since $X_{11} \sim (N-1)/r$, so it can be combined with $v_{ext}$ with the same asymptotic behavior to give a $Q/r$ term with $Q=-Z+N-1$. On the other hand,  $X_{10}$ can play a large role
\begin{align}\label{eq:X10}
X_{10}&(\bx)= \int \Psi_1^{N-1}(\bx_2 \dots\bx_N)^*  \notag \\
            & \sum_{j>1}\frac{1}{|r_j-r|} \Psi_0^{N-1}(\bx_2\dots\bx_N) d\bx_2 \dots d\bx_n
\end{align}
We are dealing with the situation that the ion ground state $\Psi_0^{N-1}$ is antisymmetric under reflection of all electronic coordinates with respect to the HNP, while $\Psi_1^{N-1}$ is symmetric. The operator in Eq.\ \eqref{eq:X10} can be split in a symmetric and an antisymmetric part with respect to reflection of the coordinates $\br_j$ in the $(xy)$ plane,
\begin{align}\label{eq:1/rasympt}
 \sum_{j>1}\frac{1}{|\br_j-\br|} &= \frac{1}{2}\sum_{j>1} \left( \frac{1}{|\br_j-\br|} - \frac{1}{|\br_j-\hat{\sigma_h}\br|} \right) \notag \\
 &+\frac{1}{2}\sum_{j>1} \left( \frac{1}{|\br_j-\br|} + \frac{1}{|\br_j-\hat{\sigma_h}\br|} \right) 
\end{align}
where $\hat{\sigma_h}$  is the operator for reflection in the horizontal ($xy$) plane, $\hat{\sigma_h}(x\hat{i}+y\hat{j}+z\hat{k})=x\hat{i}+y\hat{j}-z\hat{k}$, where $\hat{i}, \hat{j}, \hat{k}$ are unit vectors along the $x, y$ and $z$ axes.  Only the antisymmetric part will survive, and one may consider the asymptotic behavior for points $\br$ far beyond the positions $\br_j$ occurring in the integral, which range over the limited domain where the ion wavefunctions $\Psi_0^{N-1}$ and $\Psi_1^{N-1}$ have appreciable values. One may deduce that the leading term in the asymptotic behavior of $X_{10}$ will be
\begin{align}\label{eq:X10asymptotic}
&X_{10}(\br \to \infty) = \frac{\cos\theta}{r^2} \int \Psi_1^{N-1}(\bx_2 \dots\bx_N)^*  \notag \\
            & \left( \sum_{j>1}z_j\right) \Psi_0^{N-1}(\bx_2\dots\bx_N) d\bx_2 \dots d\bx_n=\frac{k\cos\theta}{r^2}
 \end{align}   
This leads to a term $(k\cos\theta/r^2) d_0$ in Eq.\ \eqref{eq:d1}, where $k$ is the integral in \eqref{eq:X10asymptotic}. The symmetries with respect to the nodal plane (both $\Psi_0^{N-1}$ and the operator in the matrix element $k$ are antisymmetric with respect to the HNP, while $\Psi_1^{N-1}$ is symmetric) do not force the $k$ integral to be zero. However, in special cases $k$ will still be zero for symmetry reasons, if for instance $\Psi_1^{N-1}$, $\Psi_0^{N-1}$ and the operator $\sum_{j>1}z_j$ belong to such irreducible representations of the molecular point group that the integral is zero. (The same argument then holds for further terms in the expansion of \eqref{eq:1/rasympt} with odd powers of $z_j$.) This is not an esoteric possibility. It holds for instance for the prototype molecules ethylene and benzene. Suppose that $d_1$ also does not couple to $d_0$ indirectly (through coupling to higher Dyson orbitals which themselves might couple to $d_0$), then $d_1$ will decay in the plane as $e^{-\sqrt{2I_1}\,r_p}$ like everywhere else, since in the eigenvalue equation \eqref{eq:Dysoneqn} all the terms except those from $\nabla^2$ can be neglected in the asymptotic region compared to $I_1$. According to Eq.\ \eqref{eq:Dysonorb} the density will, for points in the HNP,  decay as $|d_1(\bx_p)|^2\sim e^{-2\sqrt{2I_1}\,r_p}$, i.e.\ different in the plane than outside the plane. We have to consider the possibility that this special case  occurs. As will be seen, it is the only case where the KS potential can have the simple, generally assumed uniform asymptotic $-1/r$ behavior. This will be discussed as Case 1 in this section, while in section \ref{sec:Case2} we will discuss Case 2 where the density has the slow decay according to $I_0$ also on the plane, either because the $k$ integral is not zero and $d_1$ couples to $d_0$, or because higher Dyson orbitals couple to $d_0$. We emphasize that it does not appear to be likely that no higher $d_i(i>1)$ would couple to $d_0$, so Case 1 with its regular $-1/r$ asymptotic KS potential must be exceptional, if it exists at all. \\
\\
Supposing then that neither $d_1$ nor the higher Dyson orbitals $d_i(i>1)$ have the slow $e^{-\sqrt{2I_0}\,r_p}$ decay, we proceed to show that a consistent picture of the density decay in the HNP can be given, with the regular $-1/r$ asymptotic behavior of the KS potential. First we consider the question if the decay $e^{-2\sqrt{2I_1}\,r_p}$ of $|d_1(\br_p)|^2$ will actually be the decay of the total density, i.e.\ whether the infinite summation over the other Dyson orbitals squared does not overrule the decay of the first term, cf.\ Eq.\ \eqref{eq:Dysonorb}. The relation between $d_0$ and the density as in Eq.\ \eqref{d0} (but in general directions) has been used by LPS \cite{LevyPerdewSahni1984} to show that indeed the decay of $|d_0(\br_p)|^2$ and the density are the same. In the HNP we now have to use the analogous equation for $d_1$, 
\begin{align}
\label{d1} 
&d_1(\bx_p)=\sqrt{n(\bx_p)} \braket{\Psi_1^{N-1}}{\Phi(2 \cdots N|\bx_p)}_{2..N}. 
\end{align}
We show that the special properties of $\Phi$ in this case afford the required relation between $|d_1(\bx_p)|^2$ and $n(\bx_p)$ for $|\br_p| \to \infty$. The conditional amplitude $\Phi$ is a normalized $(N-1)$-electron wavefunction that describes, when $|\br_p| \to \infty$, the probability distribution of the electrons that remain behind  when one electron is infinitely far away (in this case in the plane).  The integral in Eq.\ \eqref{d1} must be $ \le 1$, since $\Phi$ and $\Psi_1^{N-1}$ are both normalized. We can also show that it cannot decay to zero, but will go to a finite constant. Consider the expansion of the ground state wavefunction in the leading KS independent particle determinantal wavefunction $\Psi_{s,0}^N$, plus all its excitations (which are orthogonal to the leading term). It is elementary to show from properties of the determinant that for $\br_p$ in the HNP the conditional amplitude  $\Phi_s(2 \cdots N|\bx_p)$ of the KS determinantal wavefunction reduces to the second ion state $\Psi_{s,1}^{N-1}$ of the noninteracting ion (the determinant with a hole in HOMO$-1$) for $r_p  \to \infty$ (see Appendix \ref{app:CondAmplHNP}). Therefore the full $\Phi$ will consist in large part of this KS ion state (the contribution of the HF or KS determinant in the wavefunction is typically substantial, 80\% - 90\% is not uncommon). The overlap of the exact second ion state with the second KS ion state, $\braket{\Psi_1^{N-1}}{\Psi_{s,1}^{N-1}}$, will be a finite constant. The integral in Eq.\ \eqref{d1} therefore remains finite. We will show below that the integral is actually 1, since $\Phi$ ``collapses" to $\Psi_1^{N-1}$ for $r_p\to\infty$, but at this point the fact that the integral goes to some finite constant is sufficient to see that the exponential decay of $\sqrt{n(\bx_p)}$  must be the same as that of $d_1(\bx_p)$.   
We can thus conclude that in the present Case 1 the {\em exact} density will have a different (faster) decay in the HNP ($e^{-2\sqrt{2I_1}\,r_p}$) than in general directions ($e^{-2\sqrt{2I_0}\,r_p}$).

\subsection{Asymptotic behavior of the Kohn-Sham potential}\label{subsec:Case1asymptoticKSpot}
 What happens in this case with the KS potential for $|\br_p| \to \infty$?\\
 The argument of Ref.~\onlinecite{WuAyersYang2003} for a negative constant does not apply here since it was based on uniform decay of the density in all directions like $\text{exp}[-2\sqrt{2I_0}\,r]$ while in Case 1 the exact density has a different decay in the HNP than elsewhere. If the exact density decays like $e^{-2\sqrt{2I_1}\,r_p}$ on the HNP, the decay $\text{exp}[-2\sqrt{2(-\epsilon_{H-1}+v_s(\infty))}\,r_p]$ of the HOMO$-1$ density can represent the density decay with a simple KS potential that goes uniformly like $-1/r$ ($v_s(\infty)=0$) if $\epsilon_{H-1}=-I_1$. Since the decay of both HOMO$-1$ and HOMO are uniform (the same outside and in the plane) the problems related to the angular derivatives (section \ref{sec:angularlaplacian}) do not appear and the uniform $-1/r$ asymptotic behavior of the KS potential is consistent with the asymptotic behavior of these solutions.
 
It is interesting to observe that the few accurate  (but not exact) calculations that are available for the KS orbital energies \cite{ChongGritsenkoBaerends2002,GritsenkoBaerends2004a} show that $\epsilon_{H-1} \approx -I_1$. The equality has not been established to better than ca.\ 0.05 eV, since the KS orbital energies are only obtained to ca.\ 0.05~eV accuracy (the calculations of the ``exact'' orbital energies use KS potentials that are generated with the criterium that they reproduce accurate CI densities).  \\
 A $positive$ asymptotic constant for the KS potential in the HNP has been found in Refs.~\onlinecite{DellaSalaGoerling2002,DellaSalaGoerling2002b,KuemmelPerdew2003,KuemmelPerdew2003b} for the exact-exchange model. This does not imply that a positive constant on the plane is also present in the asymptotic behavior of the full KS potential \cite{Joubert2007, Holas2008}. Moreover, it is not clear whether this positive constant would modify the (exponential) decay of HOMO$-1$ in the nodal plane, see section \ref{sec:angularlaplacian}, but if  it did and the exponential decay of the HOMO$-1$ density would be $\text{exp}[-2\sqrt{2(-\epsilon_{H-1}+C)}\,r_p]\approx \text{exp}[-2\sqrt{2(I_1+C)}\,r_p]$, different from the exact density, this would not have any consequence. If the local potential of the noninteracting electron system is not explicitly required to reproduce the exact density, but is determined by some other criterium such as minimum EXX energy, we may encounter essential differences between the noninteracting density and the exact one. 
\\
In the remainder of this section we investigate if  this simple Case 1 situation (different decay of the exact density on HNP than elsewhere and a regular $-1/r$ asymptotics of the KS potential in all directions), is consistent with what is known analytically about $v_s$. We consider the KS potential  in the convenient form in which it can be written \cite{BuijseBaerendsSnijders1989}: 
\begin{equation}
\label{eq:vs}
v_s=v_{\rm ext}+v^{cond}+(v^{kin}-v_s^{kin})+(v^{N-1}-v_s^{N-1}).
\end{equation}
$v^{cond}$ has been defined in and below Eq.\ \ref{eq:condampl} and the definitions of $v_{kin}$ and $v^{N-1}$ appear in the second and third lines, respectively, of the expression for the effective potential  $v_{\rm eff}(\br)$ for $\sqrt{n({\bf r})}$, Eq.\ \ref{eq:densityeqn}, see  \cite{LevyPerdewSahni1984,BuijseBaerendsSnijders1989}
\begin{eqnarray}
\label{eq:veff}
v_{\rm eff}(\bx)  & = &  v^{cond}(\bx)+\frac{1}{2} \braket{\nabla_x\Phi(2 \cdots N|\bx) }{ \nabla_x\Phi(2 \cdots N|\bx)}   \nonumber \\
& + & \brakket{\Phi(2 \cdots N|\bx)}{\hat{H}^{N-1}-E^{N-1}_0}{\Phi(2 \cdots N|\bx)} \nonumber \\
		&\equiv & v^{cond}(\bx) +v^{kin}(\bx) \label{eq:veffeqn} + v^{N-1}(\bx).
\end{eqnarray}
The potentials $v_s^{kin}$ and $v_s^{N-1}$ in $v_s$  are defined by replacing the exact conditional amplitude $\Phi$ in the definitions of $v^{kin}$ and $v^{N-1}$ (see \eqref{eq:veff}) with the conditional amplitude for the KS determinantal wavefunction, $\Phi_s$. 

LPS \cite{LevyPerdewSahni1984} noted that each term in Eq.~\eqref{eq:veff} is everywhere nonnegative and should tend to zero asymptotically. In fact, $v^{cond}(\bx)$ [see Eq.~\eqref{eq:condampl}], being the repulsive Coulomb potential of a localized charge distribution of $(N-1)$ electrons, decays like $(N-1)/r$. With $v_{ext}=-Z/r$ and $Z=N$ for neutral systems, the expected $-1/r$ behavior emerges in $v_s$ and $v_{eff}$ if the remaining terms are asymptotically zero. However, in the presence of a HNP the asymptotic behavior of the other terms is more complicated. Clearly, we will find $-1/r$ behavior for $v_s$ if both $v^{kin}-v_s^{kin}$, and $v^{N-1}-v_s^{N-1}$ are asymptotically zero.\\ 
Considering first $v^{N-1}-v_s^{N-1}$, also called the response potential $v^{resp}$  \cite{BuijseBaerendsSnijders1989,BaerendsGritsenkoJPCA1997,ChongGritsenkoBaerends2002}, we note that $v^{N-1}$ is positive since in general $\Phi$ will not be the ground state wavefunction of the ion, so its expectation value will be larger than $E_0^{N-1}$. When $|\br| \to \infty$ it has been inferred that in general the conditional amplitude collapses to the ion ground state $\Psi_0^{N-1}$ \cite{KatrielDavidson1980} (when $s=\uparrow$ then $\Phi$ will collapse to the $M_S=-1/2$ state of the doublet ion), so that $v^{N-1}(|\br|\to \infty)\to 0$. But in the HNP this changes.  
By expanding the conditional amplitude $\Phi(2 \cdots N|\bx)$  in terms of the exact $(N-1)$-electron states,
\begin{equation}
 \Phi(2 \cdots N|\bx)=\sum_{i=0}^\infty \frac{d_i(\bx)}{\sqrt{n(\bx)}} \Psi^{N-1}_i(2 \cdots N),  
\end{equation}
we see that, since on the HNP $d_0=0$ and $|d_1(\bx_p)| (r_p \to \infty) \sim \sqrt{n(\bx_p)}$, while all higher $d_i$ decay a factor $r^{-L^*}$ faster, with $L^* \ge 2$,  the conditional amplitude tends asymptotically for $\bx_p \to \infty$ on the plane to the first-excited ion state, $\Phi \to \Psi_1^{N-1}$ (note that $\Phi$ is normalized for any position $\bx$).  This implies that
\begin{equation}
\label{eq:vN-1asymptotic}
	v^{N-1}(|\br_p|\to\infty)=E_1^{N-1}-E_0^{N-1}=I_1-I_0.
\end{equation}
This is a {\em positive} constant. It would appear in the asymptotics of $v_{\rm eff}$ only on the HNP. It can be shown that $v_s^{N-1}$ also goes to the constant $I_1-I_0$ for asymptotic points $\bx_p$ in the nodal plane, and therefore cancels $v^{N-1}$, see Eq.~\eqref{eq:vN-1asymptotic}.  We have already noticed that the conditional amplitude of the noninteracting KS system with determinantal ground state collapses to the second ion state of the noninteracting system for points in the HOMO nodal plane, so
 \begin{align}
 \label{eq:vsN-1}
&v_s^{N-1}(\bx_p) \notag  \\
&= \abrakket{\Phi_s(2 \dots N | \bx_p)} {H_s^{N-1}} {\Phi_s(2 \dots N | \bx_p)} - E_{s,0}^{N-1} \notag \\
&=\abrakket{\Psi_{s,1}^{N-1}(2 \dots N)} {H_s^{N-1}} {\Psi_{s,1}^{N-1}(2 \dots N)} - E_{s,0}^{N-1} \notag \\
&= E_{s,1}^{N-1} - E_{s,0}^{N-1}= \epsilon_H -\epsilon_{H-1} = I_1 - I_0 
\end{align} 
The response potential $v^{N-1}-v_s^{N-1}$ therefore goes to zero. \\
Turning next to $v_c^{kin}=v^{kin}-v_s^{kin}$, we observe that $v^{kin}$ can also be nonzero at infinity: when crossing the HNP, the asymptotic conditional amplitude changes from $\Psi_0^{N-1}$ to $\Psi_1^{N-1}$, so that the $\br$-derivative of $\Phi$ perpendicular to the plane can be nonzero on the HNP also when $|\br|\to\infty$. The behavior of $v^{kin}$ then depends on how $d_0(\br\to\br_p)$ goes to zero when approaching the nodal plane.  We note that for the determinantal wavefunction of a noninteracting system the Dyson orbitals are precisely the occupied independent particle orbitals. In the interacting system the first Dyson orbitals for primary ion states (those corresponding to a simple orbital ionization) still are very similar to the Kohn-Sham orbitals: overlaps are typically $> 0.999$ \cite{GritsenkoBraidaBaerends2003}. This agrees with our finding in this paper that when the KS HOMO is antisymmetric with respect to a plane, the corresponding Dyson orbital also is antisymmetric with respect to that plane. Let us then take as example that asymptotically, in spherical coordinates,  $d_0\sim f(\cos\theta)R(r)e^{-\sqrt{2I_0}\,r}$, with $f(0)=0$, and $f'(0)\ne0$, as would be the case for a $\pi$ orbital,  which has $fR=r\cos\theta=z$.  By writing $v^{kin}$ in the 
form~\cite{BuijseBaerendsSnijders1989,BaerendsGritsenkoJPCA1997,ChongGritsenkoBaerends2002}
\begin{equation}
\label{eq:vkin1}
	v^{kin}(\br)=\sum_{i=1}^\infty\frac{|\nabla d_i(\bx)|^2}{n(\br)}-\frac{|\nabla n(\br)|^2}{8 n(\br)^2},
\end{equation}
and using $d_1\sim fRe^{-\sqrt{2I_1}\,r}$, it is easy to see that 
\begin{equation}
v^{kin}(r_p\to\infty)\to \frac{1}{2}f'(0)^2\frac{R^2}{r^2}e^{2(\sqrt{2I_1}-\sqrt{2I_0})\,r},
\label{eq:vkinasymptotic}
\end{equation}
showing that $v^{kin}$ can go asymptotically to infinity on the HNP. [This is not detrimental for the solution of $\sqrt{n}$ with Eq.\ \ref{eq:densityeqn} since it can be shown that terms coming from $\nabla^2$ cancel this divergence of $v^{kin}$.] 
The complete kinetic term $v_c^{kin}=v^{kin}-v_s^{kin}$ can be evaluated using the expression for $v_s^{kin}$ analogous to Eq.~\eqref{eq:vkin1}, but now written for the KS wavefunction (note that the $H$ KS orbitals $\psi_i, i=1..H$, are the exact Dyson orbitals of the noninteracting KS system, which are a finite number in this case), 
\begin{equation}
\label{eq:vskin}
	v_s^{kin}(\br)=\sum_{i=1}^H\frac{|\nabla \psi_i(\bx)|^2}{n(\br)}-\frac{|\nabla n(\br)|^2}{8 n(\br)^2}.
\end{equation}
On the plane the KS HOMO $\psi_H$ has a node. It has the same behavior in the neighborhood of the nodal plane as the first Dyson orbital $d_0$, $\psi_H\sim f_H(\cos\theta)R_H(r)e^{-\sqrt{2I_0}\,r}$, with $f_H(0)=0$, and $f_H'(0)\ne0$. Note that asymptotically the density very close to the HNP is determined by its slowest decaying part $|d_0(|\br|\to\infty,z=\delta)|^2$. But in the KS representation it is determined by the HOMO, $|\psi_H(|\br|\to\infty,z=\delta)|^2$.  We therefore expect the first Dyson orbital and the KS HOMO to have identical behavior at the nodal plane, i.e.\ $f_H(0)=f(0)=0$, $f_H'(0)=f'(0)$ and $R_H(r)=R(r)$ for $r \to \infty$. For the HOMO$-1$ we have the asymptotic behavior $\psi_{H-1}\sim e^{-\sqrt{2I_1}\,r_p}$. We then obtain for the asymptotic behavior of $v_s^{kin}$ an expression analogous to Eq.~\eqref{eq:vkinasymptotic}, and 
\begin{align}
\label{eq:vckin}
v_c^{kin}(r_p\to\infty)=v^{kin}(r_p\to\infty)-v_s^{kin}(r_p\to\infty) \notag  \\
\to (f'(0)^2-f_H'(0)^2)\frac{R^2}{2\,r^2}e^{2(\sqrt{2I_1}-\sqrt{2I_0})\,r_p},
\end{align}
Divergence of $v_c^{kin}$ does not occur if, as anticipated, $f'(0)^2-f_H'(0)^2=0$, which requires perfect similarity between the first Dyson orbital and the KS HOMO at the nodal plane.\\
\\
We conclude that in Case 1,  under rather mild conditions on similar behavior of the KS 
HOMO and the first Dyson orbital $d_0$ at the HNP, the KS potential indeed has the simple, uniform $-1/r$ asymptotic behavior that is generally assumed.  We have not rigorously proven that a small positive or negative asymptotic constant cannot exist in the potential. At this point, however, we feel that postulating such a constant in the present Case 1 is not plausible and would require convincing proof.   \\

\section{Case 2: The density decay on HNP is exponentially the same as everywhere, although polynomially slower}\label{sec:Case2}
\subsection{Asymptotic behavior of the density}\label{subsec:Case2asymptoticdensity}  
We now investigate the possibility that the Dyson orbital $d_1$ inherits,  through Eq.\ \eqref{eq:Dysoneqn}, the slow decay $e^{-\sqrt{2I_0}\,r_p}$ from the first Dyson orbital $d_0$. In that case the exact density would not have slower exponential decay on the HNP than elsewhere. We have observed that the term coupling $d_1$ to $d_0$ in the eigenvalue equation \eqref{eq:d1} for $d_1$ can be written as $k\cos^2\theta/r^2$. It may happen that $k \neq 0$, which is Case 2 discussed in this section. The fact that $X_{10}d_0$
 is zero in the HNP (because $d_0$ is zero there and the prefactor as well) does not preclude coupling of $d_1$ to $d_0$.  With $d_0 \sim r\cos\theta e^{-\sqrt{2I_0}\,r}$, we have an inhomogeneous term $(k\cos^2\theta/r) e^{-\sqrt{2I_0}\,r}$ in \eqref{eq:d1}. Clearly, \eqref{eq:d1} can only be obeyed if this term is canceled by an equal term coming from $-(1/2)\nabla^2d_1$. This can be provided by a  term in $d_1$ proportional to $(k\cos^2\theta/r) e^{-\sqrt{2I_0}\,r} \equiv f(\theta) e^{-\sqrt{2I_0}\,r}/r$.  Eq.\ \eqref{eq:d1} becomes an identity in terms $e^{-\sqrt{2I_0}\,r}/r$ if
  \begin{align}
 \frac{e^{-\sqrt{2I_0}\,r}}{r}\left( -I_0f(\theta)+k\cos^2\theta \right)=-I_1f(\theta) \frac{e^{-\sqrt{2I_0}\,r}}{r}
 \end{align}
 yielding
 \begin{equation}
f(\theta)=\frac{k}{I_0-I_1} \cos^2\theta
\end{equation}
The presence of this $e^{-\sqrt{2I_0}\,r}/r$ term does not yet change the behavior in the HNP because $f(\pi/2)=0$. However, this term in turn necessitates a $e^{-\sqrt{2I_0}\,r}/r^2$ term, which again has  a zero prefactor in the HNP. But the $e^{-\sqrt{2I_0}\,r}/r$ and $e^{-\sqrt{2I_0}\,r}/r^2$ terms necessitate next a $e^{-\sqrt{2I_0}\,r}/r^3$ term, with a prefactor that does not become zero in the HNP and changes the asymptotic behavior of $d_1$ and of the density in that plane,
\begin{align}\label{eq:d1Case2}
&d_1 = e^{-\sqrt{2I_1}\,r} -\frac{k}{I_1-I_0}\cos^2\theta \frac{e^{-\sqrt{2I_0}\,r} }{r}  \\
&+ \frac{Qk}{(I_1-I_0)^2}\cos^2\theta \frac{e^{-\sqrt{2I_0}\,r} }{r^2}+C(\theta)\frac{e^{-\sqrt{2I_0}\,r} }{r^3} + \dots  \notag \\
&C(\theta) = \frac{Qk(\sqrt{2I_0}+Q)}{(I_1-I_0)^3}\cos^2\theta+\frac{k}{(I_1-I_0)^2}(2\cos^2\theta-\sin^2\theta) \notag 
\end{align}
We note that the $e^{-\sqrt{2I_0}\,r}/r^3$ term in \eqref{eq:d1Case2} has a nonzero part in the plane (the $\sin^2\theta$ term). Therefore, when $k \neq 0$, the density in the plane has the same exponential decay as elsewhere. The special circumstance of a HNP  shows up in an asymptotic decay of the density by a factor $1/r^8$ faster than the decay of the leading contribution $|d_0(r \to \infty)|^2$ in other directions.

\subsection{Asymptotic behavior of the Kohn-Sham potential}\label{subsec:Case2asymptoticKSpot}
If the density has the slow decay according to $e^{-2\sqrt{2I_0}\,r}$ on the HNP, this has significant consequences for the KS potential. Since the HOMO is zero in the HNP, the slow decay must come from HOMO$-1$. The HOMO$-1$ KS orbital has eigenvalue $\epsilon_{H-1}$ which is rather different from $-I_0$ (actually  $\approx -I_1$). In every other direction than HNP the KS potential is assumed to go asymptotically to zero like $-1/r$, so that the HOMO (and the total density dominated by $|\psi_{s,H}|^2$) will decay correctly according to its eigenvalue $\epsilon_H=-I_0$. The  HOMO$-1$ will then have a decay $e^{-\sqrt{-2\epsilon_{H-1}}\,r} \approx e^{-\sqrt{2I_1}\,r}$, different from $e^{-\sqrt{2I_0}\,r}$, in every other direction, even arbitrarily close to the HNP.  In the one-electron Schr\"odinger equation for the KS orbitals there is only a local potential, there is no coupling to other orbitals that could modify the asymptotic behavior. The asymptotic $e^{-2\sqrt{2I_0}\,r}$ density decay in HNP must come from HOMO$-1$, which then has to switch its ``fast" decay $e^{-\sqrt{-2\epsilon_{H-1}}\,r}$ outside the HNP to the ``slow" decay $e^{-\sqrt{2I_0}\,r}$ on the HNP.  

\begin{figure}
   \includegraphics[width=7cm]{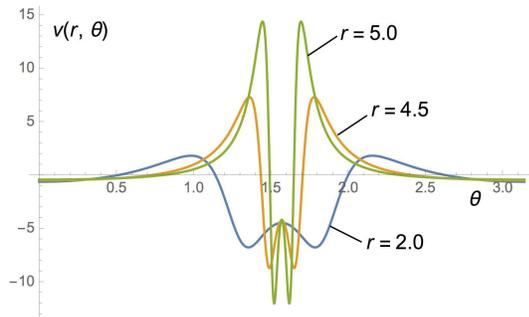}
   \caption{The potential $v(r,\theta)$ that generates the orbital with the asymptotic behavior of Eq.~\eqref{eq:psicos4}, shown as a function of the azimuthal angle $\theta=\arccos(z/r)$, for different values of $r$.  We see that although on the plane the potential goes to a negative constant, its most prominent features are actually very high positive peaks close to the plane. The region delimited by the peaks becomes very narrow (the peaks become true spikes) when $r \to \infty$, while their height grows exponentially with $r$.}
\label{fig:cos4}
\end{figure}

This raises the following question: which properties must the KS potential have in order to make  this special behavior of HOMO$-1$ possible? It is commonly assumed that the asymptotic decay is just governed by a constant in the potential. Thus, Wu et al.\ \cite{WuAyersYang2003} proposed that the decay 
$e^{-\sqrt{2I_0}\,r}$ of HOMO$-1$ in the HNP will be achieved by an appropriate constant asymptotic value of the KS potential in the HNP such that $-\sqrt{2(-\epsilon_{H-1}+v_{\rm Hxc}(r_p \to \infty))}\,r_p=-\sqrt{2I_0}\,r_p$, i.e.\ $v_{\rm Hxc}(r_p \to \infty)$ should tend to the {\em negative} constant $I_0+\epsilon_{H-1}=-(\epsilon_H-\epsilon_{H-1})$. However, the switching of asymptotic behavior generates important derivatives perpendicular to the plane, which cannot be neglected, see section \ref{sec:angularlaplacian}. Calculations on real molecular systems that are numerically exact or sufficiently accurate in the asymptotic region to demonstrate the asymptotic behavior of the KS potential are not possible. Some insight in the peculiar features that this requirement might introduce in the KS potential can be gleaned from simple analytical models. A simple analytical function which has the required decay (fast outside the plane, slow on the plane) may be written like
\begin{align}\label{eq:KSH-1}
\psi_{s,H-1}(r \to& \infty,\theta) \to \frac{e^{-\sqrt{2I_1}\,r}}  {f(\cos\theta)+ Ce^{-(\sqrt{2I_1}-\sqrt{2I_0})r}} \notag \\
                          &= \frac{1}{e^{\sqrt{2I_1}\,r}f(\cos \theta)+Ce^{-\sqrt{2I_0}\,r}}
						  \end{align}
With $f(0)=0$ we see that as long as $\theta \neq \pi/2$
\begin{equation}\label{eq:KSH-1offplane}
\psi_{s,H-1}(r \to \infty, \theta \neq \frac{\pi}{2}) \to \frac{e^{-\sqrt{2I_1}\,r}}{f(\cos\theta)}
\end{equation}
but on the plane
\begin{equation}
\psi_{s,H-1}(r \to \infty, \theta = \frac{\pi}{2}) \to \frac{1}{C}e^{-\sqrt{2I_0}\,r}
\end{equation}
The asymptotics of the KS potential can be calculated directly from the eigenvalue equation obeyed by $\psi_{s,H-1}$
\begin{equation}
v_s(\br)=\epsilon_{H-1}+\frac{1}{2}\frac{\nabla^2\psi_{s,H-1}}{\psi_{s,H-1}}
\end{equation}
Of course the results for the asymptotic behavior of $v_s$ depend on the chosen function $f(\cos\theta)$ in our example. Different choices for the function $f$ that determines the $\theta$-dependence of the switching from the decay outside the plane to the different decay on the HNP, lead to totally diferent behavior for the asymptotics of the potential in HNP, all with strong irregularities at or close to the nodal plane:
\begin{itemize}
\item for $f=\cos\theta$: $v(\br_p \to \infty) \to +\infty$ (but not applicable since $\psi_{s,H-1}$ should be symmetric with respect to the plane);
\item for $f=\cos^2\theta$: $v(\br_p \to \infty) \to -\infty$ (and has also large positive ridges around the plane, at $\theta = \pi/2 \pm \Delta$, with $\Delta$ decreasing as $r\to\infty$);
\item for  $f=\cos^4\theta$: $v(\br_p \to \infty) \to -(I_1-I_0)$ (but has large positive ridges around the plane, at $\theta = \pi/2 \pm \Delta$, with $\Delta$ decreasing as $r\to\infty$);
\item for $f=\exp(\frac{-1}{\cos^2\theta})$ : $v(\br_p \to \infty) \to -(I_1-I_0)$ (but has positive ridges around the plane, at $\theta = \pi/2 \pm \Delta$, with $\Delta$ decreasing as $r\to\infty$).
\end{itemize}
It can be seen analytically that the negative constant in the plane can arise if the first and second derivatives of $f$ (with respect to $\theta$) are zero, which is the case with the last two choices for  $f$. The behavior of $v_s$ for $f=\cos^4\theta$ is illustrated in Fig.~\ref{fig:cos4} where $v_s$ is plotted for the choice $\sqrt{2I_1}=3.0$ and $\sqrt{2I_0}=2.0$,
\begin{equation}
\psi_{s,H-1}(r,\theta)=\frac{e^{-3r}}{\cos^4\theta+e^{-2r}}
\label{eq:psicos4}
\end{equation}
We see in Fig.~\ref{fig:cos4} that although on the plane the potential goes to a negative constant, its most prominent features are actually very high positive peaks close to the plane. The region delimited by the peaks becomes very narrow (the peaks become true spikes) when $r \to \infty$, while their height grows exponentially with $r$. 

\begin{figure}
    \includegraphics[width=7cm]{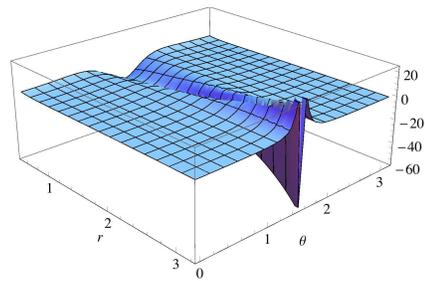}
    \caption{The potential $v(r,\theta)$ that generates the orbital with the asymptotic behavior of Eq.~\eqref{eq:KSH-1} with $f(\cos\theta)=cos^2\theta$, shown as a function of $r$ and of the azimuthal angle $\theta=\arccos(z/r)$.  We see that as $r$ increases, the potential tends exponentially fast to $-\infty$ on the nodal plane, corresponding to $\theta=\pi/2$.}
 \label{fig:cos2}
 \end{figure}

With the choice $f(\cos\theta)=\cos^2\theta$ the slowly decaying behavior of $\psi_{s,H-1}$ in the nodal pane is approached somewhat more steeply than with $f(\cos\theta)=\cos^4\theta$. The KS potential now goes to $-\infty$ on the plane for $r \to \infty$, as clearly shown in Fig.~\ref{fig:cos2}.   Notice that this diverging behavior of the potential at infinity in the HNP is compatible with a fairly regular analytic form of the eigenfunction, as in Eq.\ \eqref{eq:KSH-1}. In fact, the negative exponential divergence of the potential is canceled in the KS one-electron equation by a positive divergence from the angular part of $-(1/2)\nabla^2\psi_{s,H-1}$, as also discussed in Sec.~\ref{sec:angularlaplacian}.
 
\begin{figure}
   \includegraphics[width=7cm]{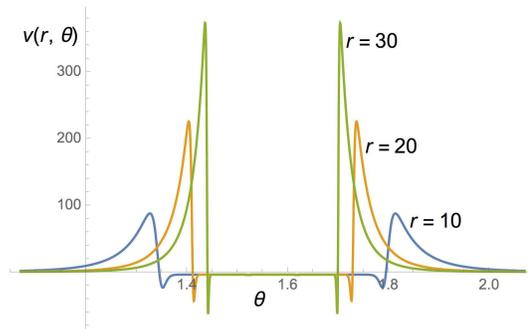}
   \caption{The potential $v(r,\theta)$ that generates the orbital with the asymptotic behavior of Eq.~\eqref{eq:KSH-1} with $f(\cos\theta)=\text{exp}(-1/\cos^2\theta)$, shown as a function of the azimuthal angle $\theta=\arccos(z/r)$, for different values of $r$.  We see that now the region close to the plane where the potential goes to a negative constant is larger than the one of Fig.~\ref{fig:cos4}. Yet, this region eventually shrinks as $r\to
   \infty$, similarly to the case of Fig.~\ref{fig:cos4}.}
\label{fig:allzeroderiv}
\end{figure}

We next choose a function that is flatter around $\theta=\pi/2$ than $\cos^4\theta$. Consider the function $f(\cos\theta)=\text{exp}(-1/\cos^2\theta)$ with all derivatives tending to zero when the plane is approached, so that the ridge in $\psi_{s,H-1}$ at any finite $r$ is rather broad, narrowing only very slowly with increasing $r$. Now at finite values of $r$ a more extended region is obtained where the potential goes to minus the constant $(I_1-I_0)$, but the pattern of diverging positive and negative peaks is again observed at both sides of the plane, see Fig.~\ref{fig:allzeroderiv}.

 We conclude that the situation that the HOMO$-1$ KS orbital has fast decay according to $e^{-\sqrt{2I_1}\,r}$ in every direction, however close to the HNP, but a slow decay $e^{-\sqrt{2I_0}\,r}$ on the plane, can be handled by special features of the KS potential. These features are however very irregular and will be very hard to represent properly in numerical approaches. Of course, these features will also affect the other KS orbitals, so the requirement that the exact total density be reproduced as the sum of squares of \textit{all} occupied KS orbitals may induce further changes in the behavior of the potential. Note that we have assumed here that the exponential decay of the HOMO is not changed by the precise behavior of the KS potential in HNP because this is a nodal plane for the HOMO. \\
 \\

\section{Summary and Conclusions}
It is known that there is an intimate relation between the asymptotic decay of the electron density and the first ionization potential of an atom or molecule. We have investigated what this relation could be when there is a nodal plane in the KS HOMO. Since the behavior of the exact density is crucial, we have invoked the expansion of the exact density in terms of squares of Dyson orbitals. The Dyson orbitals of primary ion states (those that can be associated with a simple orbital ionization) are close to KS orbitals \cite{GritsenkoBraidaBaerends2003} (the orbitals of a noninteracting electron system, like the KS system, \textit{are} the Dyson orbitals of that system). The analysis shows that one can distinguish two cases. When there would be, in the eigenvalue equation \ref{eq:Dysoneqn} obeyed by the second Dyson orbital, no coupling with the first Dyson orbital (with the same nodal plane as the KS HOMO), the second Dyson orbital would have asymptotic decay according to its eigenvalue, the second ionization potential $I_1$. The exact density would decay in the nodal plane like the square of the second Dyson orbital, $e^{-2\sqrt{2I_1}\,r}$, although everywhere else according to the first ionization energy, $e^{-2\sqrt{2I_0}\,r}$. This situation would allow for a perfectly regular KS potential which would decay uniformly (in all directions) like $-1/r$.  However, lack of coupling of the second Dyson orbital with the first requires the integral $k$ of Eq.\ \eqref{eq:X10asymptotic} to be zero. This may happen in special cases, for instance for symmetry reasons, but will not be true in general.  Then (Case 2)  the density decay will be exponentially the same everywhere, although in the nodal plane it would be polynomially slower by a factor $r^{-8}$. The KS HOMO$-1$ has to take care of this slow exponential decay in the HNP (the HOMO being zero there), while the HOMO$-1$ has decay according to its eigenvalue $\epsilon_{H-1} \approx -I_1$ in all other directions. Such behavior of the HOMO$-1$ puts very special demands on the KS potential. Apart from the negative constant (or perhaps $-\infty$) to which it should tend asymptotically in the nodal plane, it will have to have strong, asymptotically increasing, oscillations around that plane.

We should stress that many of our arguments are based on plausibility and examples, rather than on rigorous mathematical proofs. Thus, we hope that the comprehensive study presented here will trigger interest in developing a true rigorous basis for our findings.

\section*{Acknowledgments}
It is our pleasure to dedicate this paper to Andreas Savin, who always enjoyed to discover strange features in the Kohn-Sham potential and has been a pioneer in asking fundamental questions in exact DFT.\\
We thank the Netherlands Science Foundation NWO for a visitors grant for TG and a Vidi grant for PG-G, and the WCU (World Class University) program of the Korea Science and Engineering Foundation (Project No. R32-2008-000-10180-0) for support. PG-G acknowledges useful discussions with A. G\"orling and S. K\"ummel.

\appendix
\section{Collapse of the KS conditional amplitude $\Phi_s(2 \cdots N|\bx_p)$ to the first excited KS ion state for $r_p \to \infty$}
\label{app:CondAmplHNP}
For a noninteracting particle system the conditional amplitude
\begin{align}
\label{appeq:condampl}
\Phi_s(\bx_2 \dots \bx_N | \bx_1) = \frac{\Psi_{s,0}(\bx_1 \dots \bx_N)} {\sqrt{n(\bx_1)}}
\end{align}
 ``collapses" to the \textit{second} ion state for the reference point $\bx_1$ going to infinity in the HNP. 
To see this, we consider the determinantal ground state $\Psi_{s,0}$ and first let the points $\bx_1$ go to infinity in a direction where the HOMO is nonzero. The only important contribution to $\sqrt{n(\bx_1)}$ for $\bx_1 \to \infty$ will be $\phi_H(\bx_1)$. In the numerator of the conditional amplitude, Eq.\ \eqref{appeq:condampl}, we can expand the determinant. Every term where $\bx_1$ is in another orbital than $\phi_H$  will be negligible. In the remaining terms the factor $\phi_H(\bx_1)$ in the numerator cancels against the same factor in the denominator. We are left with a determinantal wavefunction with only the other orbitals, i.e. with the first ion state where the HOMO has been removed.\\
Now suppose the points $\bx_1$ are in the HOMO nodal plane. Expanding the determinant, all the terms with $\bx_1$ in the HOMO are zero. The terms with $\bx_1$ in another orbital  than the  HOMO$-1$ will be negligible.  So we retain the ones with $\bx_1$ in HOMO$-1$. The decay of the density in HNP is governed by HOMO$-1$, i.e.\ $\sqrt{n(\bx_1)}$ for $\bx_1 \to \infty$ will be $\phi_{H-1}(\bx_1)$. So we will have cancellation of $\phi_{H-1}$ on the numerator and denominator. We are left with a determinant in which $\phi_{H-1}$ has been crossed out, i.e. the conditional amplitude has collapsed to the second ion state. \\


\end{document}